\documentclass[12pt]{iopart}
\usepackage{graphicx}
\usepackage{amsfonts}
\usepackage{tikz}
\usetikzlibrary{matrix}

\begin{document}

\bibliographystyle{unsrt}

\article{}{Symmetrization of advection-diffusion operators}
\author{E. Dedits$^2$, A. C. Poje$^{1,2}$, T. Sch{\"a}fer$^{1,2}$, J. Vukadinovic$^{1,2}$}
%\address{Department of Mathematics \& CUNY Graduate Center}
\address{$^1$Department of Mathematics, College of Staten Island \\  $^2$ Physics Program at the CUNY Graduate Center}

\ead{tobias@math.csi.cuny.edu}

\begin{abstract}
  We present a new method to transform an expanded class of 
  non-selfadjoint
  advection-diffusion operators into self-adjoint operators. The
  transform is based on a combination of a point transform and Lie
  transform in conjunction with an asymptotic expansion in terms of
  the diffusivity. We illustrate the method in the context of simple
  shear flow where the expansion is exact and all
  transformation steps can be performed explicitly.
\end{abstract}
%\pacs{1315, 9440T}
\pacs{47.10A-,02.30.Jr}
\submitto{\JPA}
\maketitle

\noindent 
\section{Introduction}
In this work, we are concerned with the well-studied advection-diffusion equation in two space dimensions which we write in the following form: 
\begin{equation} \label{adv_diff_eq}
c_t + 2(u\cdot\nabla)c = \kappa \Delta c\,,
\end{equation}
where $u$ is a given incompressible velocity field, $\kappa$ is a given diffusion constant, $c=c(t,\xi)$ with $\xi = (x,y)$  is the unknown concentration function and the factor of two is included for future algebraic convenience. Most of the difficulties encountered 
in the study of this equation
arise from the fact that the advection operator $2u\cdot\nabla$ and the
diffusion operator $\kappa \Delta$ exhibit two different kinds of symmetries: the advection operator is skew-symmetric while the diffusion operator is symmetric and
hence self-adjoint. As a result, the combined advection-diffusion operator 
possesses, in general, neither symmetry. This lack of symmetry complicates 
the study of solutions of Eq.~\ref{adv_diff_eq}, especially in
the dynamically interesting case of small diffusivity $\kappa$ \cite{majda-kramer:1999}. The theory of non-self-adjoint (NSA) operators is much less developed than the theory of 
self-adjoint (SA) operators \cite{davies:2007}.  The main reason for this lies in the fact that the self-adjoint theory has a powerful tool in the spectral theorem, as well as a variety of variational methods, which can be used to obtain tight bounds on eigenvalues both theoretically and numerically.  The self-adjoint theory and its techniques have been used to great effect in quantum mechanics.  The non-self-adjoint theory, on the other hand, is much less cohesive and much more diverse.  It comprises of a variety of diverse methods with the commonality that all of them in one way or another use ideas from the analytic function theory.  The theory is still remarkably incomplete. For example, the classification of the spectra of NSA is far behind that of SA. The only exception are non-self-adjoint operators that are similar to a self-adjoint operator, so that the powerful machinery of the spectral theorem applies.  While rare for differential operators, these situations do arise. This paper is devoted to the exploration of the question when this is the case for the advection-diffusion operators.    

For a very particular class of advection-diffusion operators, 
namely those associated with irrotational (potential) velocity fields $u$, 
it is well-known that a simple point
transform  maps Eq.~\ref{adv_diff_eq} into a selfadjoint
Schr{\"o}dinger-like problem \cite{risken:1984}. 
Consequently, the transformed equation can be studied using standard 
techniques from quantum mechanics, in particular WKB approximations and 
other variational methods, giving insight into the spectral properties of the original equation, such as the
semi-classical limit.  This point-transform approach has also been used  
in the context of Fokker-Planck equations \cite{millonas-reichl:1992}. 
From both the physical and dynamical systems points of view,  cases where 
the velocity field in is time-dependent are of particular interest. 
In one spatial dimension, all flows are potential and this fact allows one 
to apply point transforms to fairly general one-dimensional non-autonomous 
convection-dominated parabolic equations. Another advantage of the transformed equation is that the 
difficulties arising from the fact that convection-dominated parabolic 
equations do not satisfy the spectral-gap condition, which is required in order for the Floquet theory of PDEs to apply.  
Using averaging techniques and inverse
scattering theory, it is possible to further transform the
non-autonomous parabolic 
equation, at least in one spatial dimension, into an 
autonomous Schr{\"o}dinger-like equation \cite{chow-lu-etal:1994}. 

Point transformations of this kind prove useful not only for linear 
autonomous and non-autonomous parabolic equations, but also for certain 
nonlinear parabolic equations.  
Point transforms have been used to circumvent the spectral-gap condition and 
prove the existence of inertial manifolds for a class of nonlinear nonlocal 
Fokker-Planck equations and for a class of viscous Burgers equations with 
low-wavenumber instability in both one and two space dimensions 
\cite{vukadinovic:2008b,vukadinovic:2008a}. For the latter, the necessary point transformation is the well-known Cole-Hopf 
transformation. The existence of inertial manifolds implies, in particular, 
that possibly very complicated global attractors can be embedded in smooth 
finite-dimensional manifolds on which the original PDE reduces to a finite 
system of ODEs. As such, the finite-dimensionality of 
the global dynamics is rigorously established.  

These remarkable results for equations featuring advection by potential flows 
naturally lead to a search for a similarity transformation which symmetrizes  
advection-differential operators involving advection by rotational flows.  
As we shall see later, there exists no simple point transform that maps 
Eq.~\ref{adv_diff_eq} into a Schr\"odinger-like equation.  
For flows with nonvanishing vorticity, we propose instead a 
combination of a point transform with an appropriate Lie-transform. As we shell see in the paper, the required Lie transform is nonlocal, and in the particular case of 
parallel flows, it is closely related to the Weierstrass integral transform, while the combined point-Lie transform is closely related to the bilateral Laplace transform. 

Our approach  is perturbative in the diffusivity and 
leads to a rather intricate system of nonlinear  `balance' equations that
the coefficients of the point and Lie transforms must satisfy. 
Obtaining and classifying solutions of these equations for general advecting fields appears 
very involved and goes beyond the scope of the
present paper.  It turns out, however, that these balance equations can be solved explicitly and easily for a class of physically interesting 
flows, such as parallel flows.  This allows us to illustrate the applicability of this approach in the context of
a simple, canonical example.
\section{Point and Lie Transforms}
We begin this section by assuming that $u$ is a given, not necessarily
incompressible flow, and illustrate how one arrives at the point (coordinate) 
transformation that symmetrizes the equation for irrotational flows, and
why the same approach fails for flows with non-vanishing curl.  For
simplicity, let us assume
that the velocity field is time independent while noting that all
results can be
extended to the non-autonomous case.  

Consider a simple point transform
of the form
\begin{equation} \label{point_trans}
c = {\mathrm{e}}^{\phi/\kappa}\,v\,.
\end{equation}
Substitution of (\ref{point_trans}) into (\ref{adv_diff_eq}) yields
the following equation for $v$:
\begin{equation}
v_t + Bv = \frac{1}{\kappa}fv+\kappa\Delta v\,.
\end{equation}
The operator $(1/\kappa) f+\kappa\Delta$ on the right hand side of the transformed equation
is again symmetric, and the potential $f$ given by
\begin{equation}\label{f}
f = \phi_x^2+\phi_y^2-2u_1\phi_x-2u_2\phi_y = |\nabla \phi|^2-2u\cdot\nabla \phi
\end{equation}
The `new' advection operator $B$  is given by
\begin{equation}
B = 2(u_1-\phi_x)\partial_x+2(u_2-\phi_y)\partial_y-\Delta\phi\,,
\end{equation}
and one can easily verify that  $B$ is also skew-symmetric. Now, we can see
directly that, if $u$ is a potential flow, we can choose
$\phi$ such that $\nabla \phi = u$, and we obtain $B=0$. Thus, we have
transformed the advection-diffusion equation
involving a skew-symmetric advection operator on the left hand side
and a symmetric diffusion operator on the right hand side into a self-adjoint 
Schr{\"o}dinger-type problem given by
\begin{equation} \label{curl-free_v}
v_t  =V v, \qquad V = \frac{1}{\kappa}f+\kappa\Delta \,.
\end{equation}
whose evolution is governed by the self-adjoint operator $V$. We are now in a situation where we the spectral theorem for (possibly unbounded) self-adjoint operators applies. We
express the solution $v(t,\xi)$ in terms of the eigenfunctions
of $V$, and obtain the solution $c(t,\xi)$ by applying the inverse point transform to this
solution. Denoting the evolution operator of $c$ as ${\mathcal{C}}$, the point transform as
${\mathcal{P}}$, and the evolution operator of $v$ as ${\mathcal{V}}$, we can illustrate 
the sequence of operations by the following diagram:
\vspace{0.2cm}
\begin{center}
\begin{tikzpicture}
\matrix (m) [matrix of math nodes,row sep=3em,column sep=4em,minimum width=2em]
{
     c(0,\xi) & v(0,\xi) \\
     c(t,\xi) & v(t,\xi) \\};
  \path[-stealth]
    (m-1-1) edge node [left] {$\mathcal{C}$} (m-2-1)
            edge node [above] {$\mathcal{P}$} (m-1-2)
    (m-2-2.west|-m-2-1) edge node [below] {${\mathcal{P}}^{-1}$} (m-2-1)
    (m-1-2) edge node [right] {$\mathcal{V}$} (m-2-2);
\end{tikzpicture}
\end{center}

In order to illustrate method described above let us consider a simple potential flow in
form $u = (u_1,u_2) = (\gamma x, -\gamma y)$, for which this flow equation 
(\ref{curl-free_v}) takes form
\begin{equation}
  \label{eq:equation_for_v}
  v_t = \Bigl(\kappa\Delta - \frac{\gamma^2}{\kappa}(x^2+y^2)\Bigr)v
\end{equation}
involving the self-adjoint Schr{\"o}dinger-operator with a quadratic potential.
The corresponding eigenfunctions involving Hermite-polynomials $H_n$ are well-known from quantum mechanics and the
solution writes as
\begin{equation}
  v = \sum_{n,m=0,1,2...} {\mathrm{e}}^{\lambda_{n,m}t}\,\alpha^{(0)}_{nm}\psi_{nm}(x,y) \label{eq:result1}
\end{equation}
where the eigenfunctions $\psi_{n,m}$ are given by
\begin{equation}
\psi_{n,m} = \sqrt{\frac{\gamma}{\kappa \pi\;2^{n+m}n!m!} }
H_n\left(\sqrt{\frac{\gamma}{\kappa}}x\right)
H_m\left(\sqrt{\frac{\gamma}{\kappa}}y\right)
\exp\Bigl(-\frac{\gamma}{2\kappa}(x^2+y^2)\Bigr) 
\end{equation}
and the discrete eigenvalues are expressed as 
\begin{equation}
\lambda_{nm} = -2\gamma(n+m+1)\,.
\end{equation}
The coefficients $\alpha^{(0)}_{nm}$ are found from
initial conditions for $v$ via projection on the corresponding eigenfunctions, hence
\begin{equation}
  \label{eq:expans_coefficients_initial}
  \alpha^{(0)}_{nm} = \int dxdy\; v_0(x,y) \psi_{nm}(x,y)\,.
\end{equation}
We can interpret the above result as follows: In the case of vanishing vorticity, a simple 
point transform (\ref{point_trans}) transforms the advection-diffusion equation into an equation involving a symmetric operator, which allows for application of the
spectral theorem. On the other hand, the above calculation shows also that, for a velocity field with non-vanishing vorticity,
a simple point transform (\ref{point_trans}) does not symmetrize the advection-diffusion equation for any choice of $\phi$. The fundamentally new idea of this work is to employ a further transform, more precisely a Lie transform,
\begin{equation} \label{lie_transform}
v = {\mathrm{e}}^{\kappa L}w, \qquad L^+=L\,.
\end{equation}
in addition to the point transform (\ref{point_trans}). The function $\phi$ and the operator $L$ are to be chosen in a way
that, at least for small diffusivities $\kappa$, the transformed problem is self-adjoint. 

Note that most of our arguments are based on formal computations in the framework of operational calculus, but they can be made more rigorous by framing the problem in an appropriate  Hilbert space of functions. One motivation for choosing a Lie transform is the fact that this
allows us to systematically construct approximate solutions in the
limit $\kappa\rightarrow 0$ by application of the Baker-Campbell-Hausdorff
formula \cite{schaefer-poje-etal:2009}.
The result, retaining terms of order ${\mathcal{O}}(\kappa^2)$, is
\begin{equation} \label{lie_expansion}
w_t + Bw + \kappa [B,L] w  =
\left(\frac{1}{\kappa}f + [f,L] + \frac{\kappa}{2}\left[[f,L],L\right]\right)w + \kappa
\Delta w  \;.
\end{equation}
Note that, at order ${\mathcal{O}}(\kappa^2)$ and higher, a variety of
terms appears, including commutators that involve not only $f$, $B$
and $L$, but also $\Delta$ and $L$. Our aim is to choose $\phi$ and
$L$ in a manner that the equation for $w$ involves only self-adjoint 
operators.  Since $B$ is skew-symmetric and $L$ is symmetric, one can easily verify that 
$[B,L]$ is symmetric. On the other hand, $[f,L]$ will be
skew-symmetric and $\left[[f,L],L\right]$ will be 
symmetric. Therefore, at leading order in the expansion using $\kappa$
as small parameter, in order to balance the skew-symmetric
operators in a manner that their influence on the evolution of $w$
vanishes, we require
\begin{equation} \label{gradient_balance}
B = [f,L]\,.
\end{equation}
Assuming that we were able to find a $\phi$ and an operator $L$ such
that the balance (\ref{gradient_balance}) of gradients is
satisfied, the resulting equation for $w$ becomes 
\begin{equation} \label{evolution_w}
w_t = \kappa A w, \qquad A = -\frac{1}{2}[B,L]+\Delta+\frac{1}{\kappa^2}f\,,
\end{equation}
where we have omitted higher order terms. If we denote the evolution operator of
$w$ by ${\mathcal{W}}$ and the Lie-transform by ${\mathcal{L}}$, we can illustrate the method by the following diagram:

\vspace{0.2cm}
\begin{center}
\begin{tikzpicture}
  \matrix (m) [matrix of math nodes,row sep=3em,column sep=4em,minimum width=2em]
  {
     c(0,\xi) & v(0,\xi) & w(0,\xi) \\
     c(t,\xi) & v(t,\xi) & w(t,\xi) \\};
  \path[-stealth]
    (m-1-1) edge node [left] {$\mathcal{C}$} (m-2-1)
            edge node [above] {$\mathcal{P}$} (m-1-2)
    (m-1-2.east|-m-1-3) edge node [above] {$\mathcal{L}$} (m-1-3)
    (m-2-2.west|-m-2-1) edge node [below] {${\mathcal{P}}^{-1}$} (m-2-1)
    (m-2-3.west|-m-2-2) edge node [below] {${\mathcal{L}}^{-1}$} (m-2-2)
    (m-1-3) edge node [right] {$\mathcal{W}$} (m-2-3);
\end{tikzpicture}
\end{center}

Note that, in general, the
commutator $[B,L]$ will not vanish and $A$ will not be a
Schr{\"o}dinger operator. By construction, however, $[B,L]$ is
symmetric and hence $A$ will be symmetric. In order to satisfy
symmetry requirements at higher orders in the expansion in $\kappa$,
it is also possible to expand $\phi$ and $L$ in an asymptotic series as
\begin{displaymath}
\phi = \sum_k \kappa^k \phi_k, \qquad L = \sum_k \kappa^k L_k
\end{displaymath}
in order to derive a hierarchy of equations for $\phi_k$ and $L_k$. In
this paper, however, we will only consider expansions up to
$\mathcal{O}(\kappa^2)$.

So far, the only requirement we have imposed on the operator $L$ is
its symmetry, $L=L^+$. Note that, at this stage, we have not imposed any conditions on the function
$\phi$, so that there is great freedom in
the choice of $L$ and $\phi$. The assumption $L=L^+$ can also be
modified, to e.g. $L=-L^+$, or dropped entirely by writing $L=L_s+L_a$
where $L_s$ denotes the symmetric and $L_a$ the skew-symmetric
part. This will change the condition for balancing the gradients in
(\ref{gradient_balance}). In this paper, we are restricting our
attention to the symmetric case. 

As $B$ is a first-order differential operator and $f$ is a function,
it is reasonable to search for an appropriate $L$ in the family of
symmetric second-order differential operators. Symmetry requires
$L$ to be of the form
\begin{equation} \label{L_form}
  L = c_{11}\partial_{xx}+2c_{12}\partial_{xy}+c_{22}\partial_{yy}+b_1\partial_x +b_2\partial_y
\end{equation}
with the conditions
\begin{equation}
b_1 = c_{11x}+c_{12y}, \qquad b_2 = c_{12x}+c_{22y}
\end{equation}
One major advantage of choosing a differential operator is that the
commutators are relatively easy to calculate. From
(\ref{gradient_balance}), the balance equations become 
\begin{eqnarray}
c_{11}f_x + c_{12}f_y &=&\phi_x-u_1 \label{balance_equation1} \\
c_{12}f_x + c_{22}f_y &=&\phi_y-u_2 \label{balance_equation2}\,.
\end{eqnarray}
Strictly speaking, from (\ref{gradient_balance}) we actually obtain a
third equation given by
\begin{equation} \label{balance_equation3}
Lf = \Delta \phi\,.
\end{equation}
An easy calculation, however, shows that if
(\ref{balance_equation1},\ref{balance_equation2}) are satisfied,
(\ref{balance_equation3}) will be satisfied as well, and the search for the desired transformation  reduces to solving the 
balance equations (\ref{balance_equation1},\ref{balance_equation2}) in conjunction with the equation (\ref{f}) for calculating the potential $f$ from the potential $\phi$. Note that solving
(\ref{balance_equation1},\ref{balance_equation2},\ref{f}) for a given velocity
field $u$ means finding functions $\phi$, $c_{11}$, $c_{12}$, $c_{22}$. 
While the balance equations may appear simple at first glance, they involve quadratic nonlinearities 
since $f$ depends on $\phi$ through (\ref{f}). For purposes of this paper, we restrict our investigation to
simple advecting fields where solutions are readily determined. 
\section{Simple shear flow}
In this section, we will examine transformation equations for the simplest example of 
advection by  linear, constant vorticity, shear flow.  To summarize the previous section, we have shown that, if a solution of the balance equations (\ref{balance_equation1},\ref{balance_equation2}) together with (\ref{f}) can be found, the
original advection-diffusion equation (\ref{adv_diff_eq}) can be
transformed (up to terms of order ${\mathcal{O}}(\kappa^2)$) into a
selfadjoint problem given by (\ref{evolution_w}).  Let us consider a velocity field of the form
\begin{equation}
u = \left(\begin{array}{c} \alpha y \\ \beta x\end{array}\right)\,.
\end{equation}
We assume $\alpha\neq\beta$, so that $\nabla\times u \neq
0$ is satisfied. As $u$ is formed by terms that are linear in $x$ and $y$, we
assume a quadratic form for $\phi$:
\begin{equation} \label{phi_guess}
\phi = \frac{1}{2}ax^2+cxy+\frac{1}{2}by^2\,
\end{equation}
with constants $a,b,c$ to be determined later. Then the expression for
$f$ yields
\begin{equation}
f = (a^2+c^2-2\beta c)x^2+(b^2+c^2-2\alpha c)y^2+2(ac+bc-\alpha a - \beta b)xy\,.
\end{equation}
Due to the particular structure of the velocity field, we can write
\begin{equation}
\nabla f = F \left(\begin{array}{c} x \\ y \end{array}\right), \qquad
F = \left(\begin{array}{cc} F_{11} & F_{12} \\ F_{21} & F_{22} \end{array}\right)
\end{equation}
with a symmetric matrix $F$ whose coefficient can be directly read
from the explicit formula for $f$. The system of equations
(\ref{balance_equation1},\ref{balance_equation2}) can now be written
in matrix form as
\begin{equation}
CF = M:=\left(\begin{array}{cc} a & c-\alpha \\ c-\beta & b \end{array}\right)
\end{equation}
and thus we can express $C$ in terms of $\{a,b,c\}$ together with
$\{\alpha,\beta\}$. By solving this system explicitly, we find that the requirement  $\alpha\neq\beta$, yields the condition 
\begin{equation} \label{square_condition}
c^2=ab\,.
\end{equation}
The explicit formula for $C$ under this assumption is
\begin{equation}
  2\gamma^2C = \left(\begin{array}{cc}
      \alpha^2a + \alpha\beta b - \beta c b - \alpha b c &
      (\alpha+\beta)c^2-2\alpha\beta c \\
      (\alpha+\beta)c^2-2\alpha\beta c &
      \beta^2b + \alpha\beta a - \alpha c a - \beta a c \end{array}\right)
\end{equation}
where we used the abbreviation $\gamma = \alpha a - \beta b$.
For the explicit calculation of $A$ governing the evolution of $w$, we
find that the coefficients of the commutator $[B,L]/2$ are given by the
matrix $2MC$. Writing
\begin{equation}
\Delta-\frac{1}{2}[B,L] = w_{11}\partial_{xx}+2w_{12}\partial_x\partial_y
+w_{22}\partial{yy}
\end{equation}
we can easily find all the entries of the corresponding matrix $W$. 
%
%the expression
%
%\begin{equation}
%\gamma^2W = \left(\begin{array}{cc} 
%\beta^2b^2 + \alpha^2ab - 2\alpha^2\beta c & 
%\alpha\beta^2b + \alpha^2\beta a - c\beta^2 b - c\alpha^2 a \\
%\alpha\beta^2b + \alpha^2\beta a - c\beta^2 b - c\alpha^2 a &
%\alpha^2 a + \beta^2ab - 2\alpha\beta^2 c \end{array}\right)\,.
%\end{equation}
%
Direct calculation shows that the determinant of $W$ is given by
\begin{equation} \label{det_formula}
{\mathrm{det}}(W) = -\frac{\alpha^2\beta^2}{\gamma^2}\,
\end{equation}
hence the resulting operator $A$ is either hyperbolic or parabolic,
but not elliptic. 

If one of the constants, $\alpha$ or $\beta$
vanishes, the velocity field reduces to a simple shear flow.  Without loss
of generality, let us consider
\begin{equation}
u = (\alpha y,0)\,.
\end{equation}
For this case, we find the matrix $c_{ik}$ as
\begin{equation}
c_{11}=\frac{1}{2a}-\frac{cb}{2 \alpha a^2}, \qquad c_{12}=\frac{b}{2 \alpha a}, \qquad c_{22}=-\frac{c}{2 \alpha a}
\end{equation}
and 
\begin{equation}
[B,L]=2\left(1-\frac{b}{a}\right)\partial_{xx}+4\frac{c}{a}\partial_{xy}\,.
\end{equation}
Remember that the evolution of $w$ is governed by the operator $A$
defined in (\ref{evolution_w}). For the first two terms in the definition
of $A$ we have found
\begin{equation}\label{generalized_shear}
 -\frac{1}{2}[B,L]+\Delta = \frac{b}{a}\partial_{xx}+\partial_{yy}-2\frac{c}{a}\partial_x\partial_y
\end{equation}
In this case, it follows from (\ref{det_formula}) that the evolution
of $w$ is governed by a {\em parabolic} equation. The potential $f$ is
given by
\begin{equation}
f = (a^2+c^2)x^2 +2 (ac+cb-\alpha a)xy + (c^2+b^2-2 \alpha c)y^2
\end{equation}
\section{Weierstrass and bilateral Laplace transforms}

Note that the above calculation for parallel flows $u = (\alpha y,0)$ follows without change for any single component, divergence-free
vector field. In particular, if instead of $u=(\alpha y, 0)$ one considers 
\begin{equation}
u = \left(\begin{array}{c} \alpha(y) \\ 0 \end{array}\right)
\end{equation}
where $\alpha(y)$ is any differentiable function, the balance equations are also satisfied
by the transform pair (chosing $b=c=0$ in the above)
\[
\phi = \frac{1}{2}ax^2 \, ,  \;\;\; L = \frac{1}{2a} \partial_{xx} 
\]
and the final transformed evolution is given by
\begin{equation}
w_t = \left(\kappa\partial_{yy}+\frac{ax}{\kappa}(a x -2 \alpha(y))\right)w.
\end{equation}
For this particular case, there is a remarkable relationship between the point and Lie transforms on one side, and the Weierstrass and the bilateral Laplace transforms on the other.  
Recall the definitions of  the bilateral Laplace transform (in one variable) 
\[
{\mathcal B}[f](\xi)=\int_{\mathbb{R}} f(x)e^{-\frac{a}{\kappa}\xi x}\, dx, \ f\in L_{\rm loc}^1(\mathbb{R})
\]
and the Weierstrass transform 
\[
{\mathcal W}[f](\xi)=\sqrt{\frac{a}{2\pi\kappa}}\int_{\mathbb{R}} f(t)e^{-\frac{a}{2\kappa}(\xi-x)^2}\, dx, \ f\in L_{\rm loc}^1(\mathbb{R}).
\]

Recall that, formally, 
$
e^{\kappa L}={\mathcal W}. 
$
Also,  the following relationship between the Weierstrass and the bilateral Laplace transform can be easily verified: 
\[
{\mathcal W}[f](\xi)=\sqrt{\frac{a}{2\pi\kappa}}e^{-\phi(\xi)/\kappa}{\mathcal B}[f(x)e^{-\phi(x)/\kappa}](-\xi), 
\]
or equivalently 
\[
{\mathcal B}[g](\xi)=\sqrt{\frac{2\pi\kappa}{a}}e^{\phi(\xi)/\kappa}{\mathcal W}[g(x)e^{\phi(x)/\kappa}](-\xi). 
\]
Recall also, 
\[
{\mathcal B}[f]^{(k)}(\xi)=\left(-\frac{a}{\kappa}\right)^k\int_{\mathbb{R}} x^k f(x)e^{-\frac{a}{\kappa}\xi x}\, dx. 
\]
In view of the above properties, it is easy to see why the same equation (\ref{generalized_shear}) is obtained if instead of the point-Lie transform pair if we substitute $c(s)={\mathcal B}[w](-s)$ in (\ref{adv_diff_eq}).  Let us note here that while the point-Lie transform approach is more general and applies also to cases of non-parallel flows, the bilateral Laplace transform approach can be made more rigorous.  Provided that $e^{-cx}f(x)\in L^2(\mathbb{R})$ for all $c\in\mathbb{R}$, the integral in the definition of the bilateral Laplace transform converges in the $L^2$ sense for all $\xi\in\mathbb{C}$ and defines an analytic function $F(\xi)$.  Moreover, the following inversion formula holds in $L^2$ sense: 
\[
f(x)=\lim_{R\to\infty}\frac{1}{2\pi i}\int_{c-iR}^{c+iR} F({\xi})e^{x\xi}\ d\xi, \ x\in \mathbb{R}. 
\]
Moreover, one easily obtains Parseval's relation
\[
\int_{\mathbb{R}} e^{-2cx}|f(x)|^2\ dx=\frac{1}{2\pi }\int_{\mathbb{R}} |F(c+i\tau)|^2\ d\tau.
\]

\section{Comparison of explicit solutions}
It is instructive to see in detail how the transformation works for a
particular initial condition, for which both the original equation (\ref{adv_diff_eq}) as well as the transformed equation can be solved explicitly via the Fourier transform on a finite time interval. This will allow us to compare the two solutions and verify the validity of our method.  Let us look for  the solution to the original
advection-diffusion equation (\ref{adv_diff_eq}) for the linear shear flow, $(\alpha y, 0)$,
with an initial condition given by
\begin{equation}
c(t=0,x,y)=\delta(x)\delta(y)\,.
\end{equation}
Note that, in this case, the analytical solution for $c(t,x,y)$ is
well-known and can be obtained directly from (\ref{adv_diff_eq}) using
Fourier transform and the method of characteristics.
%For later
%comparison we note that, in Fourier domain, we find
%%
%\begin{equation}
%\hat c(t,\omega,\eta) = {\mathrm{e}}^{-\kappa\left((\omega^2+\eta^2)t+2\omega\eta t^2 + \frac{4}{3}\omega^2t^3\right)}
%\end{equation}
%
%In this equation and in all that follow, $\omega$ will be the
%Fourier-conjugated variable with respect to $x$ and $\eta$ will be the
%Fourier-conjugated variable with respect to $y$. As for signs and factors,
%we will use
%
%\begin{equation}
%\hat f(\omega) = \int {\mathrm{e}}^{-i\omega x}f(x)dx, \qquad 
%f(x) = \frac{1}{2\pi}\int {\mathrm{e}}^{i\omega x}\hat f(\omega)d\omega
%\end{equation}
%%
%After inverse Fourier transform, we obtain immediately
%%
%\begin{equation} \label{solution_shear}
%c(t,x,y) = \frac{1}{4\pi\kappa t}\frac{1}{\sqrt{1+t^2/3}}{\mathrm{e}}^{-\frac{x^2+y^2(1+4t^2/3)-2xyt}
%{4\kappa(t+t^3/3)}}
%\end{equation}
%
Setting $(a,b,c)=(a,0,0)$  so 
that the condition (\ref{square_condition}) is trivially satisfied, 
the transformation of the initial conditions gives
\begin{equation}
w(0,x,y) = {\mathrm{e}}^{-\kappa L}\,v(0,x,y)={\mathrm{e}}^{-\kappa L}{\mathrm{e}}^{-\phi/\kappa}\delta(x)\delta(y)
\end{equation}
In order to make this equation meaningful, however, it seems reasonable to assume
$a<0$. In this way, $-\kappa L$ will be regularizing.  In frequency domain, we obtain then for $\hat w(0,\omega,\eta)$ obviously 
\begin{equation}
\hat w(0,\omega,\eta) = {\mathrm{e}}^{\kappa \omega^2/(2a)}
\end{equation}
which decays, if $a<0$. 
The next step consists in solving the equation for $w$ that simplifies to
\begin{equation}
w_t = \left(\kappa\partial_{yy}+\frac{1}{\kappa}(a^2x^2-2 \alpha axy)\right)w
\end{equation}
Using a Fourier transform only in $y$ and characteristics, we find the
explicit solution of this equation as
\begin{equation}
  \tilde w(t,x,\eta) = {\mathrm{e}}^{-\kappa\eta^2t + a^2x^2t/\kappa 
+2 \alpha aix\eta t^2 + 4 \alpha^2 a^2x^2t^3/(3\kappa)}\tilde w(0,x,\eta-2aixt/\kappa)
\end{equation}
Setting $\beta = -\kappa/(2a)>0$, we can write the initial condition
for $\tilde w$ as
\begin{equation}
  \tilde w(0,x,\eta) = \frac{1}{2\sqrt{\pi\beta}}\,{\mathrm{e}}^{-x^2/(4\beta)}
\end{equation}
and obtain the explicit formula for the Fourier transform of the solution, $\hat w =
\hat w(t,\omega,\eta)$, as
\begin{equation} \label{fourier_w_solution}
  \hat w(t,\omega,\eta) = \frac{1}{2\sqrt{\beta \gamma}}\,{\mathrm{e}}^{-(\omega-2 \alpha a\eta t^2)^2/(4\gamma)}{\mathrm{e}}^{-\kappa\eta^2t}
\end{equation}
with the abbreviation
\begin{equation}
  \gamma = \frac{-a}{2 \kappa}\left( 
1 + 2 a t \left(1 + \frac{2}{3} \alpha^2 t^2 \right) \right) \,.
\end{equation}
Note here that the Fourier transform method we employ here is valid only as long as  $\gamma>0$, i.e., on a finite time interval. After that, the Fourier transform ceases to exist, and the method leading to the explicit solution breaks down.  We can interpret this as a loss of regularity in the frequency domain in this particular step of the method, which will actually be gained back in the subsequent step of the transformation.  Moreover, $\gamma>0$ is only possible 
given our initial assumption of $a<0$.  We now transform back from $w$ to $v$ in Fourier space via
\begin{equation} \label{solution_v}
\hat v(t,\omega,\eta) =\hat w(t,\omega,\eta){\mathrm{e}}^{-\kappa \omega^2/(2a)}
\end{equation} 
In order to obtain the solution $c(t,x,y)$, we only need to transform
the solution (\ref{solution_v}) back to $(x,y)$-space and multiply by
$\exp(ax^2/(2\kappa))$. It is precisely in this multiplication, where we
get back the necessary regularity (remember $a<0$). Moreover, using
the explicit formulas for Gaussian integrals, one can easily check
that, at the end of the calculation, the parameter $a$ drops out, as
it should since $a$ was only introduced in the transformation and not
present in the original advection-diffusion equation.

For the simple example of spatially homogeneous vorticity, 
$L$ and $[B,L]$ are differential operators
with constant coefficients and all possible higher order terms in
(\ref{lie_expansion}) vanish. Therefore, in this particular case, the
transformation is exact and holds for any value of $\kappa$.

\section{Spectral representation}

In the previous example, we solved the equation for $w$ using
characteristics in order to obtain the solution
(\ref{fourier_w_solution}). In this section we show how to explicitly
use a spectral decomposition with the eigenfunctions of the operator
governing the evolution of $w$ in order to solve this equation.
Consider 
\begin{equation}
A = \partial_{yy}-\mu(x)y + \rho(x), 
\qquad \mu(x) = \frac{2\alpha ax}{\kappa^2}, 
\qquad \rho(x) = \frac{a^2x^2}{\kappa^2} \,.
\end{equation}
The (generalized) eigenfunctions of $A$ can be found by direct
calculation as
\begin{equation}
\phi_{\lambda,\tilde x} (x,y)= \mu(\tilde x)^{-1/6}{\mathrm{Ai}}
\left(\mu(\tilde x)^{1/3}y+\mu(\tilde x)^{-2/3}(\lambda-\rho(
\tilde x)\right)\delta(x-\tilde x)\,.
\end{equation}
Using the orthogonality of the Airy functions given by
\begin{displaymath}
\int_{\mathbb{R}}{\mathrm{Ai}}(t+s){\mathrm{Ai}}(t+s')\,dt = \delta(s-s')
\end{displaymath}
it is easy to show that the $(\phi_{\lambda,\tilde x})$ are
orthonormal. Setting $\tau = \kappa t$, we can write the solution to 
the equation for $w$ using these eigenfunctions as
\begin{equation}
w(\tau,x,y) = \int_{{\mathbb{R}}^2}a_{\lambda, \tilde x}(\tau)\,\phi_{\lambda,\tilde x}(x,y)\, d\tilde x d\lambda\, ,
\end{equation}
and the evolution of the expansion coefficients will be simply given by
\begin{equation}
\frac{d}{d\tau} \,a_{\lambda, \tilde x} = \lambda a_{\lambda, \tilde x}
\end{equation}
as the operator $A$ is diagonal with respect to its eigenfunctions. For our 
case in which the initial condition is given by
\begin{displaymath}
w(0,x,y) = \sqrt{\frac{-a}{2\pi\kappa}}\,{\mathrm{e}}^{ax^2/(2\kappa)}\delta(y)
\end{displaymath}
we find immediately the spectral representation of the solution $w$ as
\begin{eqnarray} 
w(\tau,x,y) &=& \sqrt{\frac{-a}{2\pi\kappa}} \int_{\mathbb{R}}
{\mathrm{e}}^{\lambda\tau}\,{\mathrm{e}}^{ax^2/(2\kappa)}
\mu(x)^{-1/3}{\mathrm{Ai}}\left(\mu(x)^{-2/3}(\lambda-\rho(x))\right)
\nonumber \\ &&
{\mathrm{Ai}}\left(\mu(x)^{1/3}y+\mu^{-2/3}(\lambda-\rho(x))\right)
\, d\lambda \label{spec_rep}
\end{eqnarray}
Note that, using the Fourier representation of Airy-functions, it is
easy to show that the following identity for Airy-functions holds
\begin{displaymath}
\int_{\mathbb{R}} {\mathrm{e}}^{\lambda t} {\mathrm{Ai}}(\lambda)
{\mathrm{Ai}}(\lambda + \alpha)\, d\lambda = \frac{1}{\sqrt{4\pi t}}\,
{\mathrm{e}}^{t^3/3}\,{\mathrm{e}}^{-(t^2+\alpha)^2/(4t)}
\end{displaymath}
and we can use this identity to compute the integral in
(\ref{spec_rep}) to obtain again the solution
(\ref{fourier_w_solution}).

\section{Conclusion}

In this work we developed a theory on how to symmetrize advection-diffusion operators using
a combined point and Lie-transform. The symmetry conditions lead to a system of nonlinear 
equations (balance equations) which are assumed to contain essential information about the 
dynamics of the advective-diffusive system. The solution of the balance equations in general
is non-trivial. We illustrated the method using a simple shear-flow for
which the balance equations can be explicitly solved. Moreover, all steps could be carried out analytically without any approximation. Extension of this
method to more general situations including time-dependent flows and periodic boundary conditions is subject of current research.

\section*{Acknowledgments}
ACP, TS and JV were supported, in part, by  grants from the City
University of New York PSC-CUNY Research Award Program.  
Also, JV was supported in part by the NSF grant DMS-0733126 and ACP by 
ONR grant N000140410192 and TS was supported in part by the NSF grants 
DMS-0807396 and DMS-1108780.

\section*{References}

\bibliography{master}

\end{document}